\documentclass[aps,prl,twocolumn,letterpaper,superscriptaddress,showpacs]{revtex4}
\usepackage{keyval}%
\usepackage{graphicx}
\usepackage{dcolumn}
\usepackage{bm}
\usepackage{color}
\usepackage{fancyhdr}
\setkeys{Gin}{width=0.8\columnwidth,clip=true}

\newcommand{\ignore}[1]{}

\ignore{
\pagestyle{fancy} %
\pagestyle{fancyplain} %
\lhead{\large arXiv:cond-mat/0509075} %
\chead{\sl to appear in PHYSICAL REVIEW LETTERS\vspace{-2pt}}
\rhead{2 Sep 2005} %
\cfoot{\sc\thepage} %
\lfoot{} %
\rfoot{} %
}

\begin{document}

\title{Orbital ordering in LaMnO$_3$: Electron-lattice versus electron-electron interactions}

\author{Wei-Guo Yin}
\affiliation{Physics Department, Brookhaven National Laboratory, Upton, NY 11973} %
\author{Dmitri Volja}
\author{Wei Ku}
\affiliation{Physics Department, Brookhaven National Laboratory, Upton, NY 11973} %
\affiliation{Physics Department, State University of New York, Stony Brook, NY 11790} %

\date{Received September 2, 2005}

\begin{abstract}
The relative importance of electron-lattice (e-l) and
electron-electron (e-e) interactions in ordering orbitals in
LaMnO$_3$ is systematically examined within the LDA+$U$
approximation of density functional theory. A realistic effective
Hamiltonian is derived from novel Wannier state analysis of the
electronic structure. Surprisingly, e-l interaction ($\simeq 0.9$
eV) alone is found insufficient to stabilize the orbital ordered
state. On the other hand, e-e interaction ($\simeq 1.7$ eV) not only
induces orbital ordering, but also greatly facilitates the
Jahn-Teller distortion via enhanced localization. Further
experimental means to quantify the competition between these two
mechanisms are proposed.
\end{abstract}
\maketitle

Study of perovskite manganites has been one of the main focuses of
recent research in condensed matter physics, not only because of
their great potentials in technological applications related to
colossal magnetoresistance (CMR) in La$_{1-x}$Ca$_x$MnO$_3$, but
also because these strongly correlated electron materials (SCEM) are
ideally instrumental to the understanding of the complex interplay
of the charge, spin, orbital, and lattice degrees of freedom that
leads to abundant fascinating phenomena, including long range order
in all the above channels \cite{CMR:dagotto}.

The unusual orbital degree of freedom in the manganites, which is
the focus of our study, originates from the singly occupied
degenerate $e_g$ states ($d_{z^2}$ and $d_{x^2-y^2}$) of the
Mn$^{3+}$ $3d$ electrons in the high-spin configuration ($t_{2g}^3
e_g^1$) due to the ligand-field splitting and strong Hund's
coupling. This orbital degeneracy makes the Mn$^{3+}$ ion
Jahn-Teller (JT) active: the degeneracy can be split via biaxial
distortion of the surrounding oxygen octahedron.

Currently, one of the critical questions on the manganites is the
interplay of electron-lattice (e-l) and $e_g$ electron-electron
(e-e) interactions
\cite{CMR:hotta,yin:prl01,lamno3:bala_00,OO:mostovoy,lamno3:ahn,lamno3:okamoto,lamno3:tyer}.
Indeed, carrier mobility \cite{yin:prl01} and magnetism
\cite{CMR:hotta,OO:mostovoy}, both essential to the unresolved
mechanism of CMR in the manganites
\cite{CMR:millis_98_nature,CMR:varma,CMR:alexandrov,CMR:ramakrishnan},
respond to these two interactions in a substantially different
manner (despite some other aspects \cite{CMR:hotta} reacting
similarly). However, even for the simplest parent compound,
LaMnO$_3$ (which presents prototype orbital order (OO) and strongest
JT-distortion in the family), various spectral measurements to date
\cite{lamno3:murakami_hill,lamno3:grenier,lamno3:kovaleva,lamno3:rauer,lamno3:kruger,lamno3:saitoh1\ignore{,lamno3:quijada,lamno3:jung,lamno3:tobe,lamno3:gruninger}}
have left two possible mechanisms of the OO in dispute: the
cooperative JT e-l effect
\cite{JT:kanamori\ignore{lamno3:millis_CJT}} and the e-e
superexchange effect \cite{OO:kugel_73}. To our knowledge, there is
no clear experimental evidence to support one over the other, due to
lack of ``signatures'' in distinguishing these two mechanisms. This
leads to great confusions in the field: for example, whether a new
type of elementary excitations called orbiton has been observed in
Raman scattering spectroscopy
\cite{lamno3:kruger,lamno3:saitoh1\ignore{,lamno3:gruninger,lamno3:brink}},
and more generally which interaction dominates localization of the
$e_g$ electrons and thus facilitates the CMR effects upon doping
LaMnO$_3$ \cite{CMR:millis_98_nature,CMR:varma}. It is thus
important and timely to discern the real roles of e-l and e-e
interactions.

In this Letter, the electronic structure of the prototype LaMnO$_3$
is systematically analyzed, aiming to quantify the relative
importance of e-e and e-l interactions in ordering the orbitals. A
realistic effective Hamiltonian for the low energy $e_g$ states is
derived from a novel Wannier function-based scheme for general SCEM,
with the key effective e-e interaction $U_\mathrm{eff}\simeq 1.7$
eV, JT splitting $\Delta_\mathrm{JT}\simeq 0.9$ eV, and
octahedral-tilting induced tetragonal crystal field $E_z\simeq 0.12$
eV. Surprisingly, e-l interaction alone is found insufficient to
stabilize OO. On the other hand, e-e interaction not only induces
OO, but also greatly facilitates the JT distortion by strongly
localizing the electrons. The present results provide new insight
into OO in LaMnO$_3$, and place stringent constraints on any
realistic theories of excitations and CMR in the manganites.
Furthermore, our analysis indicates certain \emph{competition}
between mechanisms, allowing \emph{direct} experimental
determination of their relative strengths.

Aiming at a careful determination of the relevant mechanisms without
assuming a particular dominant interaction, the following novel
three-step scheme designed for general SCEM is employed, with each
step leading to additional insights into essential interactions and
more accurate evaluation of their strengths: (i) A systematic
analysis of the electronic structure within the LDA+$U$ method
\cite{DFT:anisimov_97,note:ldau} against various constraints (e.g.:
lattice distortion), giving a rough estimate of the relevant e-l and
e-e effects; (ii) Construction of the low-energy Wannier states
\cite{wannier:ku,lamno3:grenier}, leading to signatures of
competition between e-l and e-e interactions, additional tetragonal
field effects, and a local representation of the LDA+$U$
Hamiltonian, $H^{\mathrm{LDA}+U}$; (iii) Determination of the
effective many-body Hamiltonian via a self-consistent mapping to
$H^{\mathrm{LDA}+U}$, providing a quantitative evaluation of the
interactions and deep microscopic insights.

\begin{figure}[t]
\includegraphics*{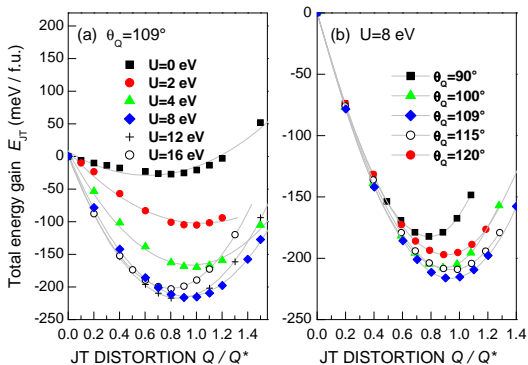}%
\caption{\label{fig:energy}%
Total energy gain per formula unit via JT distortion of the MnO$_6$
octahedra, expressed in $\mathbf{Q}_i\equiv (Q^z_i,Q^x_i) =({\sqrt 2
}(l-s),\sqrt {2/3} (2m-l-s))$, with $l$, $m$, and $s$ being the
long, medium, and short Mn-O bond lengths, respectively. For
cooperative JT distortions in LaMnO$_3$, $\mathbf{Q}_i=(Q\cos \theta
_Q ,e^{i \mathbf{q}\cdot\mathbf{R}_i} Q\sin \theta _Q )$ at the
$i$-th Mn site located at $\mathbf{R}_i$; in low-temperature phases,
$Q=Q^*\equiv 0.4$ {\AA}, $\theta^{}_Q=\theta^*_Q\equiv 109^\circ$,
and $\mathbf{q}=(\pi,\pi,0)$
\cite{\ignore{lamno3:rodriguez-carvajal,}lamno3:chatterji}.}
\end{figure}

(i) \textit{Analysis of the LDA+U results.} The results of our
systematic study of the electronic structure of LaMnO$_3$ are
summarized in Fig.~\ref{fig:energy}, where the total energy gain per
formula unit (f.u.), $E_\mathrm{JT}$, as a function of cooperative
JT distortion vectors $\mathbf{Q}_i$ (defined in the caption) is
shown for a wide range of $U$ \cite{note:ldau,note:wien}. Notice
that for realistic $U=8$ eV, the crystal structure is stabilized at
$Q^*=0.4$ {\AA} [Fig.~\ref{fig:energy}(a)] and
$\theta_Q^*=109^\circ$ [Fig.~\ref{fig:energy}(b)], in excellent
agreement with experiments
\cite{\ignore{lamno3:rodriguez-carvajal,}lamno3:chatterji},
supporting the good quality of the LDA+$U$ approximation for this
system.

Surprisingly, e-l interaction alone is found \emph{insufficient} to
stabilize the orbital ordered insulating phase: for $U=0$ eV, the
system stabilizes in a metallic state at $Q=0.8Q^*$ with small
$E_\mathrm{JT}=-27$ meV, despite being weakly insulating at $Q=Q^*$
with a small gap of 0.1 eV \cite{lamno3:pickett}. On the other hand,
$E_\mathrm{JT}$ dramatically increases in magnitude to $-215$ meV as
$U$ increases to $8$ eV, indicating that \emph{the electron
localization induced by the e-e interaction greatly facilitates the
JT instability}---in fact, OO can be stabilized even without JT
distortion \cite{DFT:anisimov_97\ignore{lamno3:medvedeva}}. Indeed,
as shown in Fig.~\ref{fig:energy}(a), fitting the data points to the
JT picture ($E_\mathrm{JT}\simeq-
\frac{1}{2}gQ+\frac{1}{2}KQ^2$~\cite{JT:kanamori}) is excellent for
$U>4$ eV but unsatisfactory for small $U$, suggesting that only with
$U>4$ eV the $e_g$ electrons are well localized as assumed in the JT
picture. For realistic $U=8$ eV, the JT splitting ($\sim
4E_\mathrm{JT}$ at $Q^*$) is thus $\Delta_\mathrm{JT}\sim 0.9$ eV,
comparable to $\sim 0.8$ eV extrapolated from spectral ellipsometry
\cite{lamno3:kovaleva,lamno3:rauer}.

It is extremely important to distinguish carefully the JT distortion
from other lattice distortions, namely octahedral tilting and its
associated octahedral distortion of GdFeO$_3$-type. The latter is
found negligible in LaMnO$_3$, as our study of LaFeO$_3$ (a
JT-inactive counterpart of LaMnO$_3$ \cite{struct:lufaso}) produces
negligible octahedral distortion with total energy gain of merely
$-3$ meV. In contrast, octahedral tilting is found to give large
energy gain of $-$403 meV/f.u. Clearly, a correct assessment of
$E_{\text{JT}}$ should exclude such a JT-distortion-unrelated
contribution. Therefore, our systematic study was performed with the
experimental octahedral tilting angle
($\phi_\mathrm{tilt}$=$16^\circ$, nearly unchanged upon JT
distortion \cite{lamno3:chatterji,struct:lufaso}).

\begin{figure}[t]
\includegraphics[width=\columnwidth,clip=true]{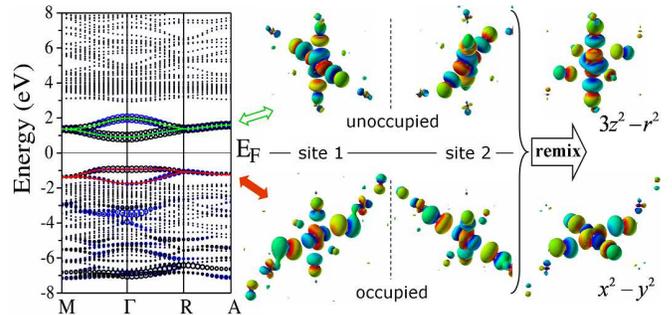}
\caption{\label{fig:BS}%
Left panel: LDA+$U$ (8 eV) band structure of LaMnO$_3$ in the real
crystal structure, with circles reflecting the weights of the Mn
$d_{z^2}$ (black) and $d_{x^2-y^2}$ (blue) symmetries.
$M$=$(\pi,0,0)$, $\Gamma$=$(0,0,0)$, $R$=$(\pi/2,\pi/2,\pi/2)$, and
$A$=$(\pi,0,\pi/2)$. Solid green and red lines show the dispersions
of OO-relevant spin-majority $e_g$ Wannier states, illustrated in
right panels (see text).}%
\vspace{-0.1cm}
\end{figure}

In addition to localizing $e_g$ states, e-e interaction plays other
crucial roles, as clearly demonstrated in the LDA+$U$ (8 eV) band
structure (Fig.~\ref{fig:BS}, left panel): The $\sim 2.6$ eV
splitting between the spin-majority occupied and unoccupied $e_g$
bands near the Fermi energy is too large to be accounted for with
the estimated JT splitting ($\sim 0.9$ eV), indicating an effective
on-site repulsion $U_\mathrm{eff} \sim 1.7$ eV. Note that the
$U_\mathrm{eff}$ relevant to OO is to be distinguished from the
``bare'' $U$ acting between atomic Mn $3d$ states, as
$U_\mathrm{eff}$ acts only between the low-energy $e_g$ Wannier
states (WSs, discussed below), and thus includes effects of
additional screening and slight delocalization via hybridization
that weaken the ``bare'' repulsion. Also note that the considerable
amount of $e_g$-character near the bottom of the oxygen $2p$ bands
($[-8,-6]$ eV) is not very relevant to OO, as such feature has its
origin in strong hybridization with the oxygen $2p$ orbitals,
independent of ordering of the orbitals. Though it does imply the
charge-transfer nature of the manganites in general.

(ii) \textit{Construction of WSs.} To proceed with more quantitative
evaluation of the above effects and to identify other relevant
mechanisms, a well-defined local (Wannier) representation of the
\emph{low-energy} $e_g$ states is necessary for further theoretical
formulation. To this end, our previously developed energy-resolved
symmetry-specific WS construction \cite{wannier:ku,lamno3:grenier}
is extended to allow mixed symmetry with a constraint search for
maximal localization \cite{note:ku}. The resulting orbital ordered
occupied ([-2.5,0] eV) and unoccupied WSs ([0,2.5] eV) are
illustrated in Fig.~\ref{fig:BS} (middle panel), from which the
staggered ordering of the orbitals is apparent, as well as the
considerable weight at the oxygen sites due to strong $p$-$d$
hybridization. Also given in Fig.~\ref{fig:BS} are band dispersions
(solid lines) corresponding to these WSs, obtained by diagonalizing
the LDA+$U$ Hamiltonian in Wanner representation,
\begin{equation}
H^{\mathrm{LDA}+U}_{jm^\prime, im}=\langle jm^\prime|
H^{\mathrm{LDA+}U} | im \rangle, \label{eq:HKS}
\end{equation}
where $| im \rangle$ denote the $m$-th WS at site $i$. For
convenient theoretical formulation, a second set of WSs with pure
$d_{3z^2-r^2}$ and $d_{x^2-y^2}$ symmetries are built from unitary
transformation (`remixing', c.f. Fig.~\ref{fig:BS}) of the first set
of WSs. The resulting ``conventional'' WSs form a realistic basis
for our theoretical modeling below.

An important observation emerges from the occupied WSs in
Fig.~\ref{fig:BS}: They do not possess the $d_{3x^2-r^2}$ and
$d_{3y^2-r^2}$ symmetry expected from the JT picture
\cite{JT:kanamori}. Indeed, the occupied states $|occ. \rangle =
\cos \frac{\theta_T}{2} |3z^{2}-r^{2} \rangle \pm \sin
\frac{\theta_T}{2} |{x^2-y^2} \rangle$ do not follow $\theta_T =
\theta_Q = 120^\circ$ from the JT model \cite{JT:kanamori}. As it
will become more apparent with the realistic Hamiltonian below,
while e-e and e-l interactions induce cooperatively the staggered OO
in this system, different mechanisms actually compete in determining
the `orbital mixing angle', $\theta_T$. Specifically, the purely
electronic (superexchange) mechanism \cite{OO:kugel_73} favors
$\theta_T=90^\circ$ for the cubic perovskite structure
\cite{lamno3:bala_00,yin:prl01}, which was confirmed by our LDA$+U$
calculations (not shown). This competition makes $\theta_T$ a
\emph{sensitive measure} of the relative importance of leading
mechanisms for OO.

\begin{figure}[t]
\includegraphics*{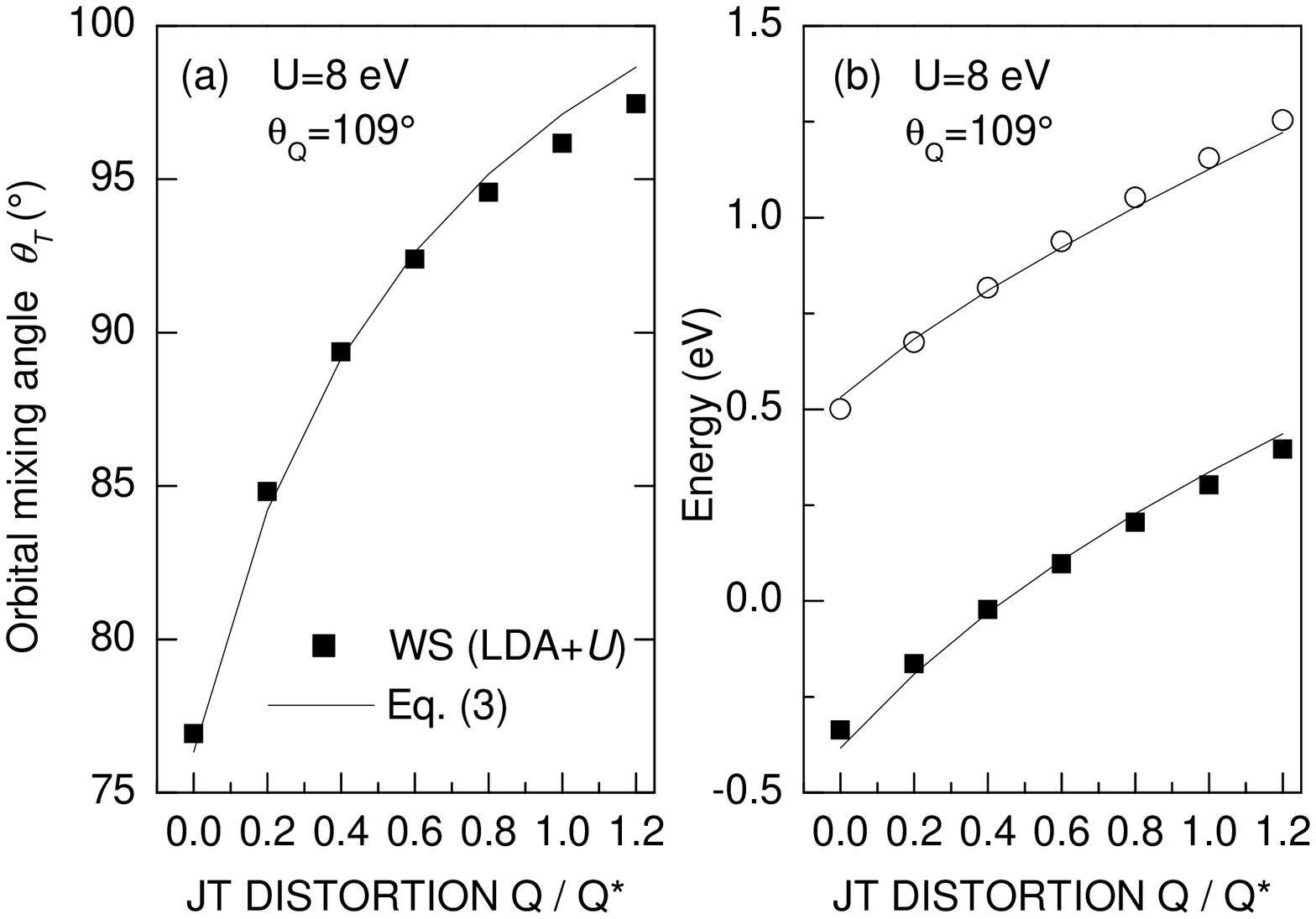}%
\caption{\label{fig:para}%
(a) $\theta_T$ as a function of JT distortion. (b) Comparison of
results from Eq.~(\ref{eq:HF}) (lines) and LDA+$U$:
$H^{\mathrm{LDA}+U}_{i\uparrow, i\uparrow} -
H^{\mathrm{LDA}+U}_{i\downarrow, i\downarrow}$ (squares) and
$H^{\mathrm{LDA}+U}_{i\uparrow, i\downarrow}$ (circles). \ignore{,
and $E_\mathrm{JT}$ (diamonds). $t=-0.6$ eV, $U_\mathrm{eff}=1.72$
eV, $gQ^*=0.87$ eV, $K=2.75$ eV/{\AA}$^2$, and $E_z=0.12$ eV are
used.}}
\end{figure}

An additional relevant mechanism can now be clearly observed from
Fig.~\ref{fig:para}(a), in which $\theta_T$, evaluated from the
above transformation between the above two sets of WSs, is plotted
against various magnitude of JT distortion $Q$ with experimental
$\theta_Q$. In the absence of the JT effect ($Q = 0$), $\theta_T =
76^\circ$ is much smaller than $90^\circ$ expected from the
superexchange effect \cite{lamno3:bala_00,yin:prl01}. This reflects
the importance of the tetragonal crystal field, $E_z$, yielded
mainly from the octahedral-titling in the real structure
\cite{lamno3:bala_00}, as $E_z$ favors $\theta_T=0^\circ$ instead.
Notice also from Fig.~\ref{fig:para}(a) that $\theta_T$ grows as $Q$
increases, reflecting a sizeable e-l interaction. Nevertheless, even
at optimal (experimental) $\mathbf{Q}$, $\theta_T$ is still way
below $\theta_Q$, confirming that the JT effect is far from being
dominant.

(iii) \textit{Mapping the Effective Hamiltonian.} Based on the above
systematic analysis, the concrete physics of the low-energy
spin-majority $e_g$ states can be described by an effective
Hamiltonian including $U_\mathrm{eff}$, $\Delta_\mathrm{JT}$, and
$E_z$:
\begin{eqnarray}
H^\mathrm{eff}&=&\sum_{\left\langle {ij} \right\rangle \gamma
\gamma^\prime} {t_{ij}^{\gamma \gamma^\prime}
d_{j\gamma^\prime}^\dag d_{i\gamma}^{}} + U_\mathrm{eff}\sum_{i}
{n_{i \uparrow } n_{i
\downarrow } } \nonumber \\
&-& g\sum_i {\mathbf{T}_i \cdot \mathbf {Q}_i } - E_z \sum_{i}
{T^z_i} + K(Q),\label{eq:model}
\end{eqnarray}
where $\gamma$ and $\gamma '$ refer to the conventional WSs,
denoting $|\uparrow\rangle=|3z^{2}-r^{2}\rangle$ and
$|\downarrow\rangle=|y^{2}-x^{2}\rangle$. $\mathbf {T}_i =
(T^z_i,T^x_i)$ with $T_{{\bf i}}^{z}=(d_{{\bf i}\uparrow
}^{\dagger }d_{{\bf i}\uparrow }^{{}}-d_{{\bf i}%
\downarrow }^{\dagger }d_{{\bf i}\downarrow }^{{}})/2$ and $T_{%
{\bf i}}^{x}=(d_{{\bf i}\uparrow }^{\dagger }d_{%
{\bf i}\downarrow }^{{}}+d_{{\bf i}\downarrow }^{\dagger }%
d_{{\bf i}\uparrow }^{{}})/2$ are pseudo-spin operators. The
in-plane nearest neighbor hopping integrals are $t_{ij}^{\uparrow
\uparrow }=3t/4$, $t_{ij}^{\downarrow \downarrow }=t/4$, and
$t_{ij}^{\uparrow \downarrow }=t_{ij}^{\downarrow \uparrow }=\mp
\sqrt{3}t/4$ (the $\mp $ sign distinguishes hopping along the $x$
and $y$ directions), while out-of-plane hoppings are strongly
suppressed in the A-type antiferromagnetic ground state due to the
double-exchange effect \cite{CMR:hotta,yin:prl01,lamno3:bala_00}.
$g$ is the JT coupling constant and $K(Q)=\frac{k}{2}\sum_i
{\mathbf{Q}_i  \cdot \mathbf {Q}_i}+\cdots$ consists of lattice
energies.

To properly map out these parameters from the LDA+$U$ results, an
unambiguous approach is developed in this study. Employing the fact
that strong local e-e interaction is approximated in LDA+$U$ in an
effective HF manner \cite{DFT:anisimov_97}, a proper connection can
be made on the same WS basis by matching
$H^{\mathrm{LDA+}U}_{j\gamma^\prime,i\gamma}$ with the
\emph{self-consistent} HF expression of $H^\mathrm{eff}$:
\begin{equation}
\ignore{H_\mathrm{HF} = }\sum_{\left\langle {ij} \right\rangle
\gamma \gamma^\prime} {t_{ij}^{\gamma \gamma^\prime}
d_{j\gamma^\prime }^\dag d_{i\gamma}^{}} -\sum_i {\mathbf{T}_i \cdot
\mathbf {B}_i } + E_0. \label{eq:HF}
\end{equation}
Here the effective field $\mathbf{B}_i = (B^z_i,B^x_i)$, where
$B_i^z = 2U_\mathrm{eff}\langle T_i^z \rangle + E_z + gQ_i^z$ and $
B_i^x = 2U_\mathrm{eff}\langle T_i^x \rangle + gQ_i^x$. $\langle
T_i^z \rangle = T^z $ and $\langle T_i^x \rangle = e^{i
\mathbf{q}\cdot\mathbf{R}_i} T^x $ with $\mathbf{q}=(\pi,\pi,0)$ are
the components of the ``pseudo-spin density wave'' mean field
$\langle \mathbf{T}_i \rangle$, which must be determined
self-consistently and is \emph{not} necessarily parallel to
$\mathbf{Q}_i$. $E_0=K(Q)+U_\mathrm{eff}|\langle \mathbf{T}_i
\rangle|^2 + U_\mathrm{eff}/4$. As shown in Fig.~\ref{fig:para}, an
excellent mapping results from $t=-0.6$ eV, $U_\mathrm{eff}=1.72$
eV, $\Delta_\mathrm{JT}=gQ^*=0.87$ eV, and $E_z=0.12$ eV.
\emph{Remarkably}, even $\theta_T= \tan^{-1} T^x / T^z$ obtained
from the effective Hamiltonian also agrees well with the LDA+$U$
results. That is, the effective Hamiltonian accurately reproduce not
only all the LDA$+U$ energies, but also the wavefunctions.

Further insight into the relative importance of e-e and e-l
interactions now can be obtained by examining the contribution of
each term of the effective Hamiltonian, Eq.~(\ref{eq:model}), to the
HF total energy, as shown in Table~\ref{table:energy}. First,
consider $\Delta_{0}$ (first row), the energy gain purely due to the
OO formation (finite $\langle \mathbf{T}_i \rangle$) in the absence
of JT distortion. Consistent with the above LDA+$U$ results, e-e
interaction alone ($-173$ meV) overwhelms the kinetic energy cost
($132$ meV) and induces OO with the total energy gain of $-50$ meV.
More intriguingly, the additional energy gain via cooperative JT
distortion $\mathbf{Q}^*$, referred to as $\Delta_{Q^*}$ (second
row), consists of considerable further contribution from e-e
interaction ($-129$ meV). Without the e-e interaction, the JT
coupling ($-356$ meV) alone is insufficient to overcome the energy
cost of lattice (96 meV) and kinetic (290 meV) energy. That is, e-e
interaction, whose contribution is enhanced due to increased
$\langle \mathbf{T}_i \rangle$, serves as a \emph{hidden driving
force} for octahedral distortion. Overall, $U_\mathrm{eff}$ is more
important to OO in LaMnO$_3$ than the JT coupling, while their
effects on cooperative octahedral distortion are comparable.

\begin{table}[t]
\caption{\label{table:energy}%
Contribution of the energy terms in Eq.~(\ref{eq:model}) to
$\Delta_0$ and $\Delta_{Q^*}$ (see text), in unit of meV per
formula unit.}
\begin{ruledtabular}
\begin{tabular}{lccccccccc}
& Total & $U_\mathrm{eff}$ & $g$ & $K$ & $E_z$ &  $t$ \\
\hline %
$\Delta_{0}$ &  $-50$ & $-173$ & 0 & 0 & $-9$ & 132 \\
$\Delta_{Q^*}$ & $-215$ & $-129$ & $-356$ & 96 & 16 & 158 \\
\end{tabular}
\end{ruledtabular}
\end{table}

The present results place stringent constraints on any realistic
theories of the manganites and interpretations of their excitation
spectra, for example, current controversy on the recently observed
$120-160$ meV Raman shifts of incident photons resonant at
$E_\mathrm{res}\simeq 2$ eV \cite{lamno3:saitoh1,lamno3:kruger}. On
the one hand, the excitations were interpreted as orbital waves, or
`orbitons', derived from the pseudo-spin superexchange model in the
large $U_\mathrm{eff}$ limit of Eq.~(\ref{eq:model})
\cite{OO:kugel_73}. With the orbiton spectrum gap $\sim
2.5J_\mathrm{orb}+\Delta_\mathrm{JT}$
\cite{lamno3:brink,lamno3:bala_00,yin:prl01} and superexchange
coupling constant $J_\mathrm{orb} \simeq 40-50$ meV ($\propto
t^2/U_\mathrm{eff}$), this scenario requires small
$\Delta_\mathrm{JT}\leq 50$ meV
\cite{lamno3:okamoto,lamno3:saitoh1,\ignore{lamno3:saitoh2,}lamno3:brink}.
On the other hand, in the JT scenario the Raman shifts were
attributed to two-phonon processes (single phonon frequency $\sim
60-80$ meV) induced by a phonon assisted \emph{on}-site $d^4_i \to
d^4_i$ transition, requiring large $\Delta_\mathrm{JT}\simeq
E_\mathrm{res}\simeq 2$ eV
\cite{lamno3:kruger,\ignore{lamno3:gruninger,}lamno3:allen_exciton\ignore{,lamno3:perebeinos_raman}}.%
\ignore{however, the two-phonon signal would be expected too small
\cite{lamno3:saitoh1}.} In contrast, our quantitative results
($\Delta_\mathrm{JT}\simeq 0.9$ eV and $U_\mathrm{eff}\simeq 1.7$
eV) show that the spin-majority $e_g$ states are in the intermediate
e-e interaction regime with comparable JT coupling.
Thus, a more reasonable picture would be two-phonon processes
mediated by optically active \emph{inter}-site
$d^4_id^4_j \to d^3_id^5_j$ transition with
$E_\mathrm{res} \simeq 2$ eV \cite{lamno3:grenier,lamno3:kovaleva}.
\emph{Direct experimental verification} of our results includes
probing, e.g., on-site $d$-$d$ transition ($\sim 0.9$ eV) via
inelastic X-ray scattering, or $\theta_T$ via nuclear magnetic
resonance.

In summary, we have quantified the relative importance of e-e and
e-l interactions in ordering orbitals in LaMnO$_3$ using a new
theoretical approach. A realistic effective Hamiltonian resulting
from this quantitative method reproduces consistently both LDA+$U$
energies and wavefunctions. Intermediate e-e interaction
($U_\mathrm{eff}\simeq 1.7$ eV) is found to play a crucial role in
inducing OO and localizing the electrons, which in turn enhances the
JT interaction ($\Delta_\mathrm{JT}\simeq 0.9$ eV) and stabilizes JT
distortion. Furthermore, a clear ``signature'' of competition
between e-e and e-l interactions is given via the orbital mixing
angle $\theta_T < \theta_Q$. Experimental means to directly clarify
the relative strengths of the leading mechanisms are suggested. The
developed general theoretical scheme can be applied to other
strongly correlated materials.

We are grateful to P. Allen, E. Dagotto, S. Grenier, J. Hill, A.J.
Millis, A. Moreo, G. Sawatzky, D. Singh, and J. Thomas for helpful
discussions. W.Y. thanks M. Lufaso and P. Woodward for providing
SPuDS \cite{struct:lufaso} and T. Chatterji for providing structural
data \cite{lamno3:chatterji}. Brookhaven National Laboratory is
supported by U.S. Department of Energy under Contract No.
DE-AC02-98CH1-886.  This work is partially supported by DOE-CMSN.


\begin{thebibliography}{29}
\bibitem[{not({\natexlab{b}})}]{note:wien}
\bibinfo{note}{All our LDA+$U$ results were obtained with the
linearized augmented plane wave method \cite{wien} for the A-type
antiferromagnetic phase with experimental volume and fixed
$J$=0.11$U$. Hypothetic crystal structures were prepared with SPuDS
\cite{struct:lufaso}.}
\end{thebibliography}

\end{document}